# Efficacy of surface error corrections to density functional theory calculations of vacancy formation energy in transition metals

**Prithwish Kumar Nandi, M C Valsakumar[‡], Sharat Chandra, H K Sahu, C S Sundar**

Materials Science Group, Indira Gandhi Centre for Atomic Research, Kalpakkam-603 102, Tamil Nadu, India

E-mail: valsa@igcar.gov.in

**Abstract.** We calculate properties like equilibrium lattice parameter, bulk modulus and monovacancy formation energy for nickel (Ni), iron (Fe) and chromium (Cr) using Kohn-Sham density functional theory (DFT). We compare relative performance of local density approximation (LDA) and generalized gradient approximation (GGA) for predicting such physical properties for these metals. We also make a relative study between two different flavors of GGA exchange correlation functional, namely, PW91 and PBE. These calculations show that there is a discrepancy between DFT calculations and experimental data. In order to understand this discrepancy in the calculation of vacancy formation energy, we introduce a correction for the surface intrinsic error corresponding to an exchange correlation functional using the scheme implemented by Mattsson et al. [Phys. Rev. B **73**, 195123 (2006)] and compare the effectiveness of the correction scheme for Al and the 3d-transition metals.

PACS number(s): 71.15.Mb, 61.72.J-, 71.20.Be, 73.90.+f

## 1. Introduction

The Kohn-Sham (KS) density functional theory [1] (DFT) based calculation forms the corner stone of *ab initio* electronic structure calculation and has been applied to problems in physics, chemistry and biology [2]. DFT focuses on quantities in real, three dimensional coordinate spaces, mainly on ground state electron density [3]. The single particle KS equations will, in principle, account for all ground state many body effects when used with the exact exchange correlation (XC) functional [3], which is not known to date. Therefore, it is clear that practical usefulness of DFT for describing ground state properties depends entirely on whether approximations for this XC functional could be found which are sufficiently simple and accurate. The simplest approximation of XC functional is the local density approximation (LDA) [1,4]. In this approximation, XC functional depends on the exchange correlation energy per particle of a uniform electron gas of a given density. This prescription is exact for a uniform electron gas and a priori expected to be fairly accurate for systems having a slow variation of electronic density on the scales of local Fermi wavelength and Thomas Fermi wavelength [3]. LDA can fail in systems where electron-electron interaction effects are dominant [3]. An important improvement over LDA is the generalized gradient approximation (GGA) of electron density where the XC functional depends on electron density and its spatial variation [5,6,7]. But all such treatments of XC functional and its consequent improvements may be inappropriate in systems for which assumption of uniform or slowly varying electron density is inapplicable [8]. According to Kohn and Mattsson, the KS single particle wavefunction makes a transition from oscillatory to a decaying type where the electron charge density makes a sharp jump [8]. Therefore, the uniform density based assumption of DFT, breaks down in describing such cases. One such situation arises in a material with a vacancy, since it introduces a steep variation of electronic density near the vacant site [9,10]. Such electronic density gradient resembles the variation near a surface region of a material. This gives rise to a qualitative difference between the perfect bulk and a system with a vacancy. The DFT based total energy calculation of such a system leads to inaccurate estimation of vacancy formation energy [9].

According to Mattsson and Kohn [10], there are two complementary ways to improve the accuracy of a DFT based calculation of vacancy formation energy:

(1) One can continue to develop more accurate local, quasi-local or universal approximations such as the LDA, GGA and weighted density approximation all of which presume enough local resemblance with the uniform electron gas. (2) Divide the material into two regions; treat the region away from the vacancy by the usual method of DFT and the region near a vacancy by using alternative methods, like analytic formulation and Monte-Carlo techniques. Finally, these two results are integrated so that both the descriptions are well matched at the boundary.

In the present work, we adopt the ideas of the second method following Mattsson *et al.* [9,11]. To set the stage for our work, we present Mattsson's work in some detail, and then present our work on 3d transition metals. For establishing the equivalence of our calculations with earlier work [11], we repeat the calculations for Al also.

A general outline of Mattsson's scheme is as follows [9,10,11,12,13]: The surface surrounding the vacancy is approximated to represent that of a simplified model system. For this system, a surface self-energy correction (energy/unit surface area) is determined as a function of electron density. Secondly, the density of the actual system is invoked to get the model system parameters. Since, exact data for both surface exchange and surface correlation energy are available for a jellium surface [9], Mattsson *et al.* implemented a correction scheme based on jellium surface model [11]. In order to quantify the surface intrinsic error per unit area, they calculated XC surface energies ($\sigma_{XC}$) for jellium surface, for various choices of XC functional. They also calculated the most accurate XC jellium surface energies ($\sigma_{XC}^{RPA+}$) for each XC functional, using the "improved random phase approximation" (RPA+) [14]. The difference between these two surface energy terms, $\Delta\sigma_{XC} = \sigma_{XC} - \sigma_{XC}^{RPA+}$, is used as the correction for surface energies for that particular XC functional [11]. A compact parameterized form of the surface intrinsic error is given by [11]:

$$\Delta\sigma_{XC}(\tilde{r}_s) = A\tilde{r}_s^{-5/2} + B\tilde{r}_s^{-3/2} . \tag{1}$$

where, $\tilde{r}_s = \dfrac{r_s}{a_{Bohr}}$ , $a_{Bohr}$ = Bohr radius and $r_s$ = Wigner Seitz radius defined as,

$$r_s = a_{lat}\left(\frac{3}{4\pi MN_{elect}}\right)^{\frac{1}{3}} = \left(\frac{3}{4\pi \bar{n}}\right)^{1/3} ; \tag{2}$$

where, $a_{lat}$, $M$ and $N_{elect}$ represent equilibrium lattice parameter, number of atoms per unit cell and number of outermost electrons per atom respectively and $\bar{n}$ is the average valence electron density in the material. For LDA, the values of A and B were estimated to be 0.028 eV/Å$^2$ and -0.0035 eV/Å$^2$ respectively. For PW91, the values of A and B are 0.0984 eV/Å$^2$ and -0.0144 eV/Å$^2$ respectively while for PBE, the values of A and B are estimated to be 0.0745 eV/Å$^2$ and -0.0109 eV/Å$^2$ respectively [11].

Therefore, there are two key ingredients to this surface intrinsic correction scheme: (a) the area of the exposed surface because of the creation of a vacancy and (b) the electron density at this surface. By matching the electron density profiles of the model jellium surface and that surrounding the vacancy in Al, as calculated using DFT, Carling *et al.*[15] have estimated the radius of a vacancy in Al to be 1.2 Å. This radius, suitably scaled by the lattice parameter, is used as the size of vacancy for other metallic systems by Mattsson [9] and we have also followed the same prescription. As for the electron density, Mattsson *et al.* have used the average bulk electron density, $\bar{n}$, defined as $MN_{elect}/a_{lat}^3$, and showed that this works reasonably well for Al, Pt, Pd and Mo [9,11]. In our work we have used Mattsson's prescription of using the average bulk electron density. In addition, we have used an alternative prescription (to be discussed later) which is shown to give better agreement with experiment.

In this paper we discuss results for vacancy formation energy ($E_f^v$) in Ni, Fe and Cr. Since these three transition metals have large electron density and highly localized 3d-orbitals, we expect to reveal large discrepancies between DFT calculations and experimental data [16,17]. These 3d systems are of technological importance and it is important to have an accurate estimate of $E_f^v$. Therefore, it is of interest to see the efficacy of Mattsson's correction method for these 3d systems.

**2. Details of calculation**

We perform the DFT calculations using VASP [18,19,20] (Vienna *Ab initio* Simulation Package) code, using plane-wave basis set. In the present calculations, we use projector augmented wave [21] (PAW) formalism based pseudopotentials (PPs). For PAW PPs we use PBE [7], PW91 [5,6] and LDA [4] XC functionals. All the PPs are taken from the VASP PP library. We take great care in ensuring convergence of all results with respect to system size, basis sets and k-points as discussed in the Appendix. All the calculations done here are based on supercell approach. We perform the calculations with various supercell sizes to study the dependence of the results on the system sizes. We find that $5 \times 5 \times 5$ supercell for fcc Ni (125 atoms) and $4 \times 4 \times 4$ supercell for bcc Fe and Cr (128 atoms) provide convergence of the total energy per atom to less than $10^{-3}$ eV. In all these calculations, we have allowed the ionic positions, volume and shape of the supercell to relax. The relaxation of atomic positions is done with the conjugate gradient method. This minimization process is terminated when the force acting on each atom is less than $10^{-5}$ eV/Å. We perform spin polarized calculations for all the three systems. For Ni and Fe we use ferromagnetic model whereas for Cr we use a simple antiferromagnetic (AFM) model where the two sublattices have alternating spin configurations (G-type) [22]. The common settings of DFT calculations for Ni, Fe and Cr are summarized in the Appendix.

## 3. Results and discussions

Here we discuss the results of the computation described above. In section 3.1, we discuss the equilibrium lattice parameters ($a_{lat}$) and bulk modulus ($B_0$) for Ni, Fe and Cr. In section 3.2, we present the results of the calculation of vacancy formation energies for these metals. The effects of surface energy correction on $E_f^v$ are discussed in section 3.3.

*3.1. Equilibrium lattice parameter and bulk modulus*

At first we calculate bulk properties like equilibrium lattice parameter and bulk modulus of Ni, Fe and Cr and compare the results with available experimental data. The values of $a_{lat}$ (in Å) and $B_0$ (in GPa) corresponding to different PPs are tabulated in Table 1. We calculate these bulk properties from basic electronic structure computation and compare the values with experimental data with a view to validate the parameterized XC functionals; these functionals are, in turn, employed in the estimation of vacancy formation energy. To obtain equilibrium lattice parameter and $B_0$, we fit energy vs. volume data of the fully relaxed cells to the Murnaghan's equation of state [24]. For this calculation, we have used nine data points within ±5% of experimental volume. From Table 1, we notice that PBE and PW91 XC functionals make accurate estimates of equilibrium lattice parameter of Ni, whereas LDA underestimates this value by ~3% when compared with experimental value. Both PBE and PW91 underestimate $a_{lat}$ for Fe and Cr by ~1% and ~2% respectively, LDA underestimates the same by ~4%. Similarly, for bulk modulus, in case of Ni, LDA overestimates the experimental value of $B_0$ as compared with PW91 and PBE values. In case of Fe, PW91 and PBE overestimate $B_0$ by ~20% and the LDA overestimates the same by ~40-50% from the experimental data. Though for Cr, PBE gives good agreement with experiment, PW91 and LDA values differ significantly from the experimental values. In order to identify the cause of the observed differences in the value of the bulk modulus obtained from the various computations, we consider the expression for the bulk modulus at 0 K

$$B(V) = V\frac{d^2}{dV^2}\int_{E_{\min}}^{E_F(V)} ED(E,V)dE = V\left[\frac{d}{dV}\left(E_F D(E_F,V)\frac{\partial E_F}{\partial V}\right) + \left(\frac{\partial D(E,V)}{\partial V}\bigg|_{E=E_F}\right)E_F\frac{\partial E_F}{\partial V} + \int_{E_{\min}}^{E_F(V)} E\frac{\partial^2 D(E,V)}{\partial V^2}dE\right],$$

where $E_{\min}$ is the minimum of the energy of the band electrons, and the volume dependence of the Fermi energy ($E_F$) is given by the constraint $N = \int_{E_{\min}}^{E_F(V)} D(E,V)dE =$ constant. We notice that $B(V)$ depends on $\frac{\partial^2 D(E,V)}{\partial V^2}$ in the entire range of energy of the band electrons. We have plotted, in Fig. 1, the density of states (DOS) vs. energy, for a fixed lattice constant of Ni for both PAW PBE and PAW LDA. This plot shows that the valence energy spectra of Ni for PAW PBE and PAW LDA are essentially identical in nature, in agreement with a similar observation by Ruban *et al.* [17], for 3d transition metals. However, as can be seen from the inset of Fig. 1, $\frac{\partial^2 D(E,V)}{\partial V^2}$ shows substantial differences between LDA and PBE. This may account for the difference in the bulk modulus, as estimated from these two schemes.

As mentioned earlier, we have carried out spin polarized calculations for all of the three metals. Magnetic moment values per unit cell for Ni and Fe are found to be 2.52 and 4.4 $\mu_B$, respectively, which are in good agreement with experimentally available data [25]. However, for Cr, we have used a simple AFM configuration (G-type) though the magnetic ground state of Cr is controversial [26]. After spin relaxation, the eventual magnetic moment was seen to be 0.99 $\mu_B$ and 0.56 $\mu_B$ for PBE and PW91 respectively; LDA gave zero magnetic moment. This is also in gross agreement with other estimates [27,28].

*3.2. Vacancy formation energy*

The formation energy is calculated using the following formula [9]:

$$E_f^v = E(N-1,1) - \frac{N-1}{N} E(N,0) \qquad (3)$$

Here, E(N,0) represents total energy of the perfect system with N atoms of the supercell and E(N-1,1) is the energy of the system when one of the atoms is replaced by a vacancy.

The calculated vacancy formation energies and corresponding experimental and computed data available in literature are tabulated in Table II for Ni, Fe and Cr.

In case of Ni, both PBE and PW91 versions of GGA underestimate vacancy formation energy by ~20% as compared to experimental value for Ni whereas LDA underestimates this within ~6%. In case of Fe, LDA overestimates the $E_f^v$ by 13%, with a better agreement (~9%) with GGA methods. In case of Cr, while LDA and PW91 overestimates the $E_f^v$ by ~23% and ~17% respectively, a very good agreement is seen with the PBE XC functional. In Table 2, we also have mentioned results from earlier calculations. Note that for Cr, our results lead to better agreement with experiment than earlier values. According to Mattsson [9], DFT underestimates the vacancy formation energy when the effect of structural relaxation is incorporated; however, in our case we observe both positive and negative deviations of $E_f^v$ from experimental data. We also observe that for Fe and Cr, GGA improves the agreement, whereas in the case Ni, usage of GGA actually worsens the agreement with experiment. We must mention that this discrepancy is unrelated to lattice relaxations, since in all our calculations, lattice relaxations have been incorporated. We already mention in section 1 that DFT makes inaccurate estimate of $E_f^v$ since the creation of a vacancy inside a material introduces a steep variation of electronic density around the vacant site. This abrupt variation of electronic density resembles the variation of the same near a surface region. Therefore it seems logical to study whether the discrepancy between DFT and experiment in estimating $E_f^v$ is related to such surface related errors.

*3.3. Surface self-energy corrections*

As indicated in section 1, surface self-energy corrections have been calculated using the method suggested by Mattsson *et al.*[11]. In order to successfully reproduce the procedure of Mattsson *et al.*, [11] we have repeated the calculations for Al before proceeding with the computation for Ni, Fe and Cr.

We have carried out a calculation using PAW PBE potential for Al for which experimental as well as computed data are already available in the literature [11]. We obtain the values of equilibrium lattice constant and bulk modulus by fitting energy vs. volume data to Murnaghan's equation of state [24]. For this calculation, we have used nine data points within ±5% of experimental volume. In Table 3, we have tabulated our results along with experimental value as well as the values calculated by Mattsson *et al.*[11]. The correction per unit area, net correction and corrected $E_f^v$ calculated using average bulk electron density of the system as prescribed by Mattsson *et al.*[11] are tabulated in the columns labeled as "MATT". The comparison of Al data with Mattsson's work [11] is satisfactory and serves to validate our computational methods. Please note that the deviations of the uncorrected $E_f^v$ (0.62 eV) and corrected $E_f^v$ (0.77 eV) from experimental value (0.68 eV) are almost comparable and therefore it appears that the correction scheme does not work well even for Al, where the jellium based model should work better. In fact, in case of Al, this correction scheme overcorrects $E_f^v$. We may argue that the choice of average valence electron density of a material to calculate the surface intrinsic error to $E_f^v$ appears to be a too simplified approach. By making such a choice we are actually losing the information of spatial distribution of electronic charge, which may be important in deciding the charge density at the surface created around the vacancy.

Here we propose an alternative way to choose the electron density for better estimating the surface related correction to $E_f^v$. We know that whenever a vacancy is created inside a material, the atoms surrounding the vacant site relax to minimize the strain in the system. The contribution of such strain energy to $E_f^v$ has been taken into account in our calculations by incorporating lattice relaxations. Similarly electrons surrounding the vacancy region also relax to minimize the kinetic energy of the system. Electrons residing far away from the vacant site remain almost unaffected while the most significant changes occur near the vacancy site as shown in Figure 1. Therefore, the new electron density for determination of surface intrinsic error could be extracted from the detail of the electron density distributed around the vacant site and comparing it with the density in the perfect lattice. The correct electron density to be used for the surface correction scheme is calculated in the following way:

We have calculated the charge density in a spherical shell of radius $r_v$ and thickness $\Delta r$ (0.03 Å) around the vacant lattice site both for the perfect system and the system with a vacancy. The radius of the spherical shell, $r_v$, is chosen such that it corresponds to the spherical surface created around the vacancy. The difference of these densities, $\Delta \rho$, is taken as the density to calculate the surface intrinsic error for these metals. The values of $\Delta \rho$ and surface related correction per unit area, total correction and corrected vacancy formation energy values for Al are tabulated in the Table 3 in the columns with label "CW". It is seen from Table 3 that the corrected value of $E_f^v$ (0.70 eV) calculated with this new density, $\Delta \rho$, is in better agreement with experiment (0.68±0.03 eV) compared to the corresponding value (0.77 eV) obtained using the average valence electron density of the system. The advantage of using the difference of electron densities at vacancy surface between the perfect and vacancy system, $\Delta\rho$, thus becomes evident from this observation.

Therefore, we have applied this newly defined density, $\Delta\rho$, for estimating the surface intrinsic errors for Ni, Fe and Cr. We have also included the corresponding results obtained using mean valence electron density of the systems for the purpose of comparison. The uncorrected $E_f^v$ and corrected $E_f^v$ for Ni, Fe and Cr are tabulated in Table 4. The results obtained with newly defined electron density, $\Delta\rho$, are tabulated in columns labeled as "CW", whereas the data computed using mean valence electron density of the system are tabulated in the columns labeled as "MATT". The data of exposed surface area due to creation of a vacancy, average bulk electron density ($\Delta\rho$), correction for Ni, Fe and Cr are provided in a table as supplementary information. First, we describe the results obtained using the mean bulk electron density to calculate the surface intrinsic error to $E_f^v$ (tabulated in the columns labeled as "MATT" in Table 4). We observe that the corrected $E_f^v$ for Ni, Fe and Cr are not close to the corresponding experimental values, rather the correction worsens the agreement with experiment. Not only that, the agreement between $E_f^v$ values as obtained using the PBE, PW91 and LDA functionals are also poor for all the cases discussed here. Actually, after adding surface related corrections to $E_f^v$, the values obtained using PBE, PW91 and LDA should become close to each other. We also note that the deviations of the corrected $E_f^v$ of transition metals from corresponding experimental values are much more pronounced than the case of Al. This can be explained from Fig.2, where we have plotted the normalized charge density (charge density/maximum charge density) vs. r/d along the close packed direction of Al, Ni, Fe and Cr for both perfect crystal and the crystal having a vacancy. We have already discussed that because of the creation of a vacancy, a surface is being created inside a material. This surface is assumed to enclose a spherical region around the vacant site. We have calculated the radius, $r_v$, of this spherical volume for Al, Ni, Fe and Cr using Mattsson's technique [9] as described in Section 1 and seen that $|r_v/d| \approx 0.4$. This is evident from Fig. 2b that the bulk density of a jellium surface approximately at $|r_v/d| \approx 0.4$, matching the Al vacancy density profile, is approximately the average bulk valence electron density, while this is not the case for Ni, Fe and Cr. This aspect is further clarified in Fig.3. In Figs. 3a, 3b, 3c and 3d, we plot the valence electron density in the (001) plane for the perfect systems of Al, Ni, Fe and Cr. These plots show that for Ni, Fe and Cr, valence charges density is a maximum at lattice sites and depleted considerably away from the atoms. However, in Al, the valence electron density is a minimum at atomic sites and is seen to spread over the interstitial spaces. We have also plotted, in Figs. 3e, 3f, 3g and 3h, the valence electron density around the vacant lattice site in the (001) plane for Al, Ni, Fe and Cr respectively. The plot, as shown in Fig. 3e, also suggests that whenever a vacancy is introduced in Al, electrons surrounding the vacant site move towards it, where as for Ni, Fe and Cr, as observed from Figs.3f, 3g and 3h respectively, valence electrons remain concentrated near atomic cores and therefore, no significant change in electronic density ensues in and around the vacant sites. As a result, the bulk density at a jellium surface at $|r_v/d| \approx 0.4$, matching the density profiles for Ni, Fe and Cr, is much less than the mean valence electron density for each metal. This particular observation also rules out the choice of using the mean bulk electron density of a system to calculate the surface intrinsic error.

Let us now describe the data obtained using the newly defined electronic charge density, $\Delta\rho$, to calculate the surface intrinsic correction to $E_f^v$ (tabulated in columns labeled as "CW" in Table 4). Please note that the correction improves the agreement between experimental (~1.8 eV) and computed values for all the three XC functionals, among which LDA (1.7 eV) gives the best matching. Both for Fe and Cr, the corrected values of $E_f^v$ for all the XC functionals are still overestimated as compared to the experimental values, but the overestimation is less as compared to the scheme outlined by Mattsson *et al.*[11]. This observation can be justified owing to the fact that the electrons in 3d transition metals are rather localized around the atoms, and therefore, the change in electron density at the surface of the created vacancy is low. As a result, the correction is small. In fact, an agreement between the individual corrected values, for each XC functional, is an indication of how good is the assumption that the surface error is the most dominating difference between the results obtained with different XC functionals for the same system. We note that the corrected $E_f^v$ for Ni with PBE (1.58 eV), PW91 (1.58 eV) and LDA (1.7 eV) match well. We also notice that, for Fe, the agreement between the corrected $E_f^v$ with PBE (2.36 eV), PW91 (2.41 eV) and LDA (2.32 eV) are quite satisfactory. In case of Cr, though the correction to $E_f^v$ improves the agreement between PW91 (2.87 eV) and LDA (2.90 eV) values, the matching with the PBE (2.43eV) value is comparatively poorer. Here we should mention that Cr is thought to have spin-density wave (SDW) based AFM experimental ground state. Since performing a SDW calculation is computationally much more expensive, we have used a G-type AFM state for our calculation. The observed discrepancy between our DFT results for Cr from experimental data may be related to this fact. However, this is clear from the above discussion that the newly defined electronic charge density, $\Delta \rho$, for calculating the surface intrinsic correction to $E_f^v$ improves the efficacy of the surface correction model of Mattsson *et al.* [11] for Al as well as for the transition metals.

## 4. Conclusions

The detailed DFT study of bulk properties like equilibrium lattice parameter and bulk modulus for the 3d-transition metals like Ni, Fe and Cr, using some XC functionals under PAW PP based formalism, has been carried out. These calculations show that GGA gives better estimate of these equilibrium properties than LDA for these metals. Our results demonstrate that both LDA and

GGA PP based DFT calculations make inaccurate estimate for vacancy formation energy. The mismatch between reported experimental value and the computed value for Cr using PW91 and LDA XC functionals are found to be quite large. Therefore, we conclude that even the so-called simple problem of calculating vacancy formation energy is not straightforward. We have attempted to resolve this issue by incorporating surface intrinsic energy corrections to $E_f^v$ using a jellium based model developed by Mattsson *et al.* [11]. This model in its original form calculates surface intrinsic error per unit area using the average bulk density of the system as a key parameter [11]. Our present study shows that this particular choice of electronic charge density gives surface intrinsic correction which increases the disagreement with experiment instead of reducing it. Here we have pointed out the reason for failure of this particular choice of electronic charge density for calculating surface intrinsic error and outlined a more appropriate method of choosing a suitable electronic charge density to be used for estimating the surface energy correction. The surface intrinsic correction calculated using so defined electronic charge density when added to uncorrected $E_f^v$ values of Ni, Fe and Cr produce values comparable with experiments. This also improves the agreement between PBE, PW91 and LDA values, except for Cr. In fact, the results show that the strength of this surface correction scheme is not in estimating the exact values of $E_f^v$ but in the identification of the cause for the error.

However, presently there exists XC functionals like AM05 [29] and PBEsol [30] that incorporates the surface effects in a self-consistent manner in DFT calculations. The low electron density in the interstices of 3d-transition metals has some resemblance to semiconductors and for interstitial defects [31], the surface effects are likely to play more important role than in the case of vacancies. This work reports evidence for the non-equivalence of PBE and PW91 in computation of defect properties as seen in the results for Cr. We have, however, not carried out investigations to establish this quantitatively.

**Acknowledgements**

One of the authors (PKN) wants to acknowledge Dr. Ravi Chinappan and Dr. Mathi Jaya for useful discussions.

**Appendix**

Common settings for all Ni calculations: plane wave cutoffs are ~337.0 eV for PAW PBE, PAW PW91 and PAW LDA whereas the recommended cutoff energies (ENMAX) are 269.533 eV, 269.561 eV and 269.618 eV for PAW PBE, PAW PW91 and PAW LDA respectively. Augmentation used ~545 eV for PAW PBE, PAW PW91 and PAW LDA. In all calculations for Ni the numbers of k-points used are $4 \times 4 \times 4$ in the Monkhorst-Pack scheme [23]. This gives the convergence of ~$10^{-5}$ eV for the total energy per atom.

Common settings for all Fe calculations: plane wave cutoffs are 335.0 eV for PAW PBE, PAW PW91 and PAW LDA whereas the recommended cutoff energies (ENMAX) are 267.883 eV, 267.907 eV and 267.969 eV for PAW PBE, PAW PW91 and PAW LDA respectively. Augmentation used ~511.4 eV for PAW PBE, PAW PW91 and PAW LDA. In all calculations for Fe the numbers of k-points used are $5 \times 5 \times 5$ in the Monkhorst-Pack scheme [23]. This gives the convergence of ~$10^{-5}$ eV for the total energy per atom.

Common settings for all Cr calculations: plane wave cutoff s are ~350.0 eV for PAW PBE, PAW PW91 and PAW LDA whereas the recommended cutoff energies (ENMAX) are ~227 eV. Augmentation used ~402 eV for PAW PBE, PAW PW91 and PAW LDA. In all calculations for Cr the numbers of k-points used are $5 \times 5 \times 5$ in the Monkhorst-Pack scheme [23]. This gives the convergence of ~$10^{-5}$ eV for the total energy per atom.

For all calculations mentioned above the energy tolerance for electronic iterations are $10^{-6}$ eV and Fermi smearing value is 0.2 eV. All calculations are performed with "PRECISION = HIGH" in the INCAR files.

**References**

1. Hohenberg P and Kohn W 1964 Phys. Rev. B 136 B864; Kohn W and Sham L J 1965 Phys. Rev 140 A1133.
2. Gillan M J 1997 Contemp. Phys. 38 115.
3. Kohn W 1999 Rev. Modern Phys. 71 1253.
4. Ceperley D M and Alder B J 1980 Phys. Rev. Lett. 45 566.
5. Perdew J P and Wang Y 1992 Phys. Rev. B 45 13244.
6. Perdew J P, Chevary J A, Vosko S H, Jackson K A, Pederson M R, Singh D J and Fiolhais C 1992 Phys. Rev. B 46 6671; 1993 Phys. Rev. B 48 4978.
7. Perdew P, Burke K and Ernzerhof M 1996 Phys. Rev. Lett. 77 3865.
8. Kohn W and Mattsson A E 1998 Phys. Rev. Lett. 81 3487.
9. Mattsson T R and Mattsson A E 2002 Phys. Rev. B 66 214110.
10. Mattsson A E and Kohn W 2001 J. Chem. Phys. 115 3441.
11. Mattsson A E, Armiento R, Schultz P A and Mattsson T R 2006 Phys. Rev. B 73 195123.
12. Mattsson A E and Jennison D R 2002 Surf. Sci. Lett. 520 L611.
13. Mattsson A E, Schultz P A, Desjarlais M P, Mattsson T R and Leung K 2005 Modelling Simul. Mater. Sci. Eng.13 R1.
14. Yan Z, Perdew J P and Kurth S 2000 Phys. Rev. B 61 16430; Pitarke J M and Eguiluz A G 2001 Phys. Rev. B 63 045116.

15. Carling K, Wahnstrom G, Mattsson T R, Mattsson A E, Sandberg N and Grimvall G 2000 Phys. Rev. Lett. 85 3862.

16. Kurth S, Perdew J P, and Blaha P 1999 Int. J. Quantum Chem. 75 889.

17. Ruban V and Abrikosov I A 2008 Rep. Prog. Phys., 71 046501.

18. Kresse G and Hafner J 1993 Phys. Rev. B 47 558.

19. Kresse G and Hafner J 1994 Phys. Rev. B 49 14251.

20. Kresse G and Furthmuller J 1996 Phys. Rev. B 54 11169.

21. Kresse G and Joubert J 1999 Phys. Rev. B 59 1758; Blöchl P E 1994 Phys. Rev. B 50 17953.

22. *Magnetism in Condensed Matter,* Stephen Blundell (Oxford University Press 2001) pp 96-97

23. Monkhorst H J and Pack J D 1976 Phys. Rev. B 13 5188.

24. Murnaghan F D 1937 Am. J. Math. 49 235; Murnaghan F D 1944 Proc. Natl. Acad. Sci. U.S.A. 30 244.

25. Pauling L and Ewing F J 1948 Rev. Modern Phys. 20 112.

26. Fawcett E 1988 Rev. Modern Phys. 60 209.

27. Hafner R, Spisak D, Lorenz R and Hafner J 2002 Phys. Rev. B 65 184432.

28. Bihlmayer G, Asada T, and Blugel S 2000 Phys. Rev. B 62 11937.

29. Armiento R and Mattsson A E 2005 Phys. Rev. B 72 085108.

30. Perdew J P, Ruzsinszky A, Csonka G I, Vydrov O A, Scuseria G E, Constantin L A, Zhou X and Burke K 2008 Phys. Rev. Lett. 100 136406, see also supplementary information.

31. Mattsson A E, Wixom R R and Armiento R 2008 Phys. Rev. B 77 155211.

32. http://www.webelements.org

33. *Atomic Point Defects in Metals : Crystal and Solid State Physics*, edited by H. Ullmaier, Landolt-Bornstein ( Springer- Verlag, Berlin 1991 ) Vol.25, pp. 211-214.

34. Positron annihilation results quoted in Schaefer H E 1987 Phys. Status Solidi A 102 47.

35. Fu C C and Willaime F 2004 Phys. Rev. Lett. 92 175503.

36. Schultz H and Ehrhart P, *Atomic Defects in Solids*, edited by Ullmaier H, Landolt-Börnstein, New Series Group III ( Springer, Berlin, 1991).

37. Korhonen T, Puska M J and Nieminen R M 1995 Phys. Rev. B 51 9526.

38. Megchiche E H, Pérusin S, Barthelat J C and Mijoule C 2006 Phys. Rev. B 74 064111.

39. Olsson P, Domain C and Wallenius J 2007 Phys. Rev. B 75 014110.

Figure Captions:

Figure1. Density of states (DOS) plot of Ni calculated with experimental lattice constant (2.867 Å) for spin up configuration. The features of the DOS for both LDA and PBE are essentially identical. In inset, we have plotted the second derivative $\partial^2 D(E,V)/\partial V^2$ of the density of states D(E,V) as a function of energy. This plot shows that though the DOS spectra of the PBE and LDA look similar, the quantity $\partial^2 D(E,V)/\partial V^2$, given by the two methods differ significantly leading to different values for the bulk modulus.

Figure 2. One dimensional plots of normalized charge density vs. r/d (a) for perfect lattice and (b) for the lattice having a vacancy. In both cases normalized charge density is defined as a ratio of charge density to maximum charge density for the same metal. Here r is distance and d is nearest neighbor distance along closed packed direction. It is to note that, the valence electron density for Al at atomic sites is very less contrary to Ni, Fe and Cr. Not only that, owing to the free electron like character, the disturbance in electron density profile spreads beyond first neighbor for Al where as for the 3d metals, the electron density remains almost unchanged even in the first neighbor distance. The highly localized 3d electrons in these transition metals are mainly responsible for this. The dotted lines in Fig. 2b indicate the approximate position of the vacancy surface.

Figure 3. Contour plots of normalized electronic charge density in (a) the (001) plane of Al, (b) (001) plane of Ni, (c) (001) plane of Fe and (d) (001) plane of Cr for perfect lattice structures. Contour plots of electronic charge density in (e) the (001) plane of Al around the vacancy, (f) (001) plane of Fe around the vacancy, (g) (001) plane of Fe around the vacancy and (h) (001) plane of Cr around the vacancy. From these figures, it is clear that whenever vacancy is created in Al, electrons from surrounding move towards it, resembling free electron like character. But for the transition metals, electrons are highly localized around the atomic sites.

Table Captions:

Table 1. The computed DFT values of equilibrium lattice parameters and bulk moduli of Ni, Fe and Cr. The numbers are calculated using various flavors of pseudopotentials: PAW PBE, PAW PW91 and PAW LDA. The computed values are compared with experimental values.

Table 2. Vacancy formation energies for Ni, Fe and Cr are calculated using PAW (PBE, PW91, LDA) pseudopotentials by DFT. Calculated values are compared with experimental data as well as other computed data.

Table 3. The computed DFT values of equilibrium lattice parameter, bulk modulus, vacancy formation energy, corrected vacancy formation energy are calculated using PAW PBE pseudopotential. The numbers are compared with experimental values [15] as well as data as calculated by Mattsson et al [11].

Table 4. Corrected values of vacancy formation energies are compared with experimental values. Corrected values using average bulk valence electron density of the material and using $\Delta\rho$ are tabulated in columns labeled as "MATT" and "CW" respectively.

Supplementary Information:
The computed values of exposed surface area as well as corresponding surface corrections are calculated. Wigner-Seitz radius ($r_s$) are calculated using Equation 2. Values using average bulk valence electron density of the material and using $\Delta\rho$ are tabulated in columns labeled as "MATT" and "CW" respectively.

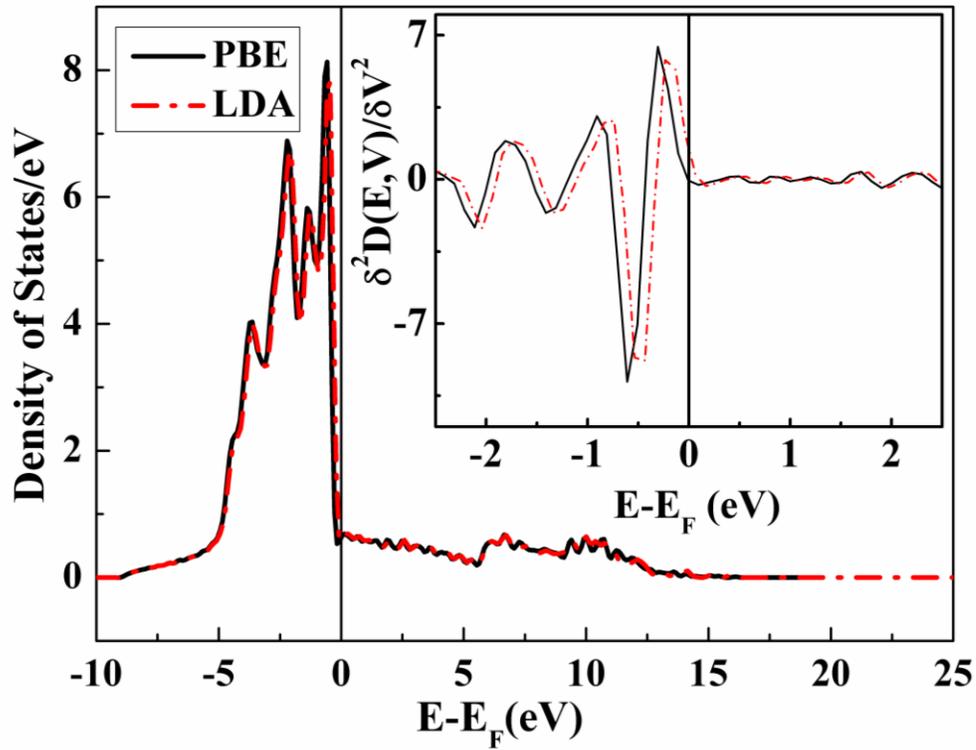

**Figure1.** Density of states (DOS) plot of Ni calculated with experimental lattice constant (2.867 Å) for spin up configuration. The features of the DOS for both LDA and PBE are essentially identical. In inset, we have plotted the second derivative $\partial^2 D(E,V)/\partial V^2$ of the density of states D(E,V) as a function of energy. This plot shows that though the DOS spectra of the PBE and LDA look similar, the quantity $\partial^2 D(E,V)/\partial V^2$, given by the two methods differ significantly leading to different values for the bulk modulus.

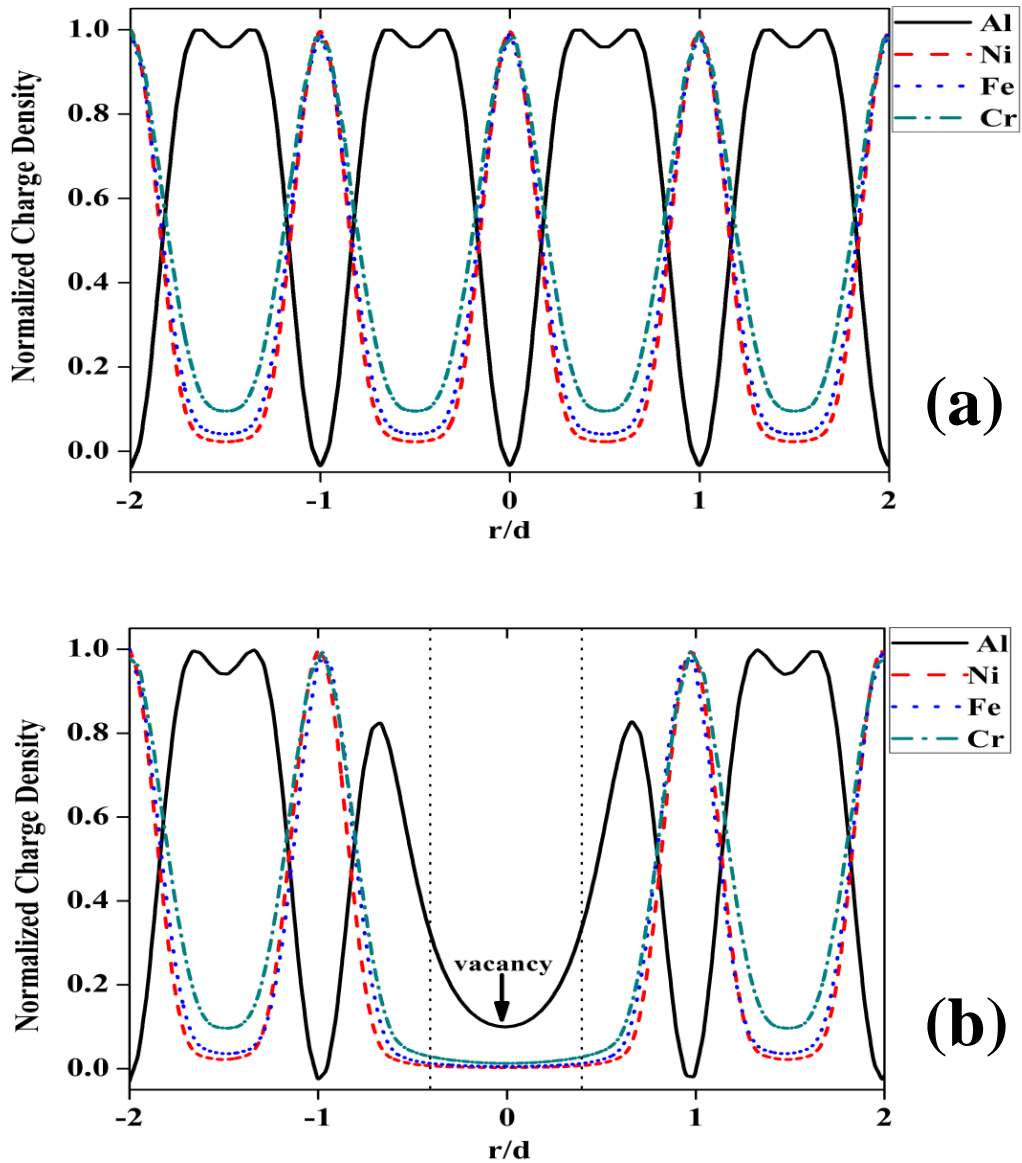

Figure 2. One dimensional plots of normalized charge density vs. r/d (a) for perfect lattice and (b) for the lattice having a vacancy. In both cases normalized charge density is defined as a ratio of charge density to maximum charge density for the same metal. Here r is distance and d is nearest neighbor distance along closed packed direction. It is to note that, the valence electron density for Al at atomic sites is very less contrary to Ni, Fe and Cr. Not only that, owing to the free electron like character, the disturbance in electron density profile spreads beyond first neighbor for Al where as for the 3d metals, the electron density remains almost unchanged even in the first neighbor distance. The highly localized 3d electrons in these transition metals are mainly responsible for this. The dotted lines in Fig. 2b indicate the approximate position of the vacancy surface.

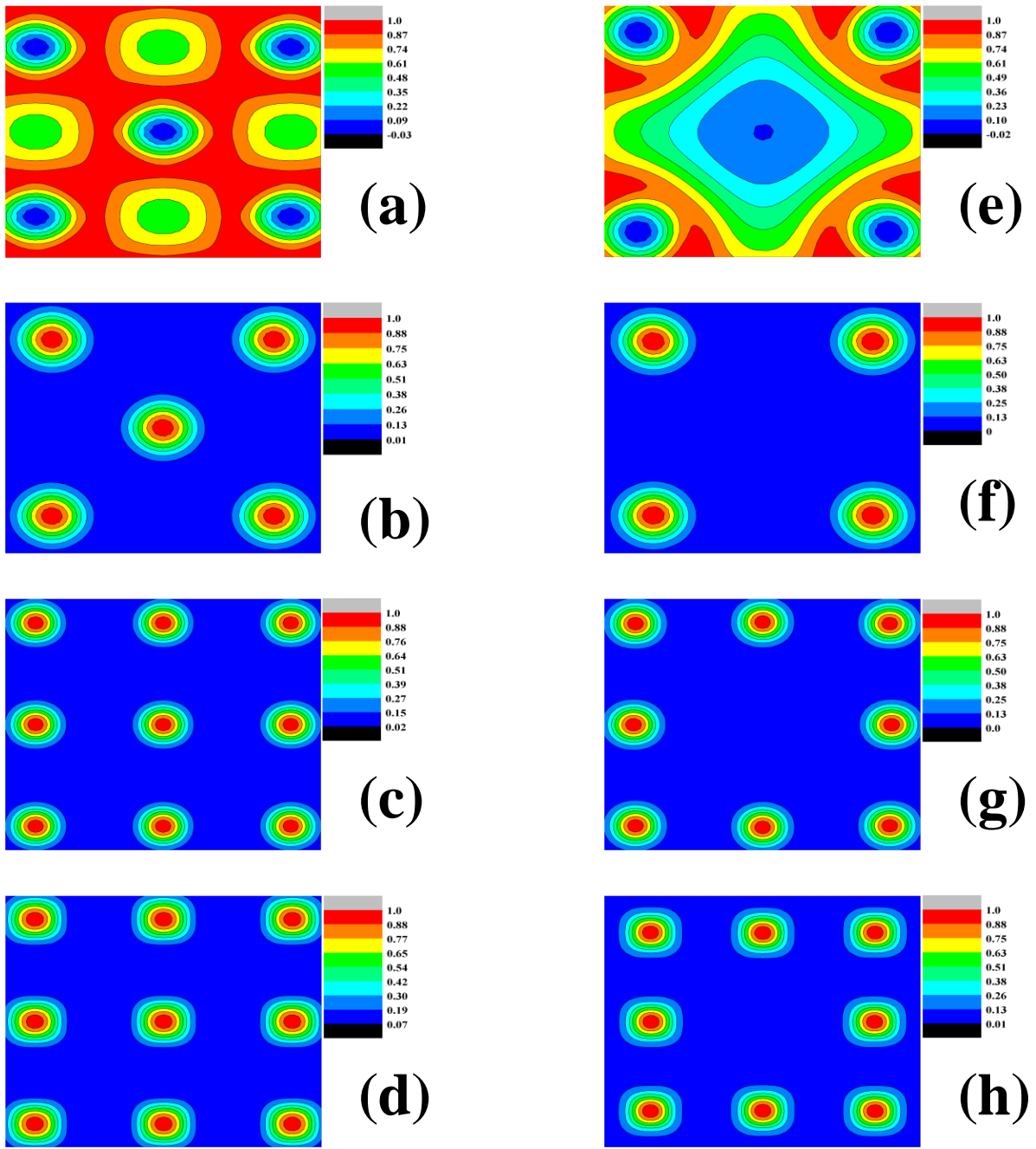

**Figure 3.** Contour plots of normalized electronic charge density in (a) the (001) plane of Al, (b) (001) plane of Ni, (c) (001) plane of Fe and (d) (001) plane of Cr for perfect lattice structures. Contour plots of electronic charge density in (e) the (001) plane of Al around the vacancy, (f) (001) plane of Fe around the vacancy, (g) (001) plane of Fe around the vacancy and (h) (001) plane of Cr around the vacancy. From these figures, it is clear that whenever vacancy is created in Al, electrons from surrounding move towards it, resembling free electron like character. But for the transition metals, electrons are highly localized around the atomic sites.

**Table 1.** The computed DFT values of equilibrium lattice parameters and bulk moduli of Ni, Fe and Cr. The numbers are calculated using various flavors of pseudopotentials: PAW PBE, PAW PW91 and PAW LDA. The computed values are compared with experimental values.

| Metals | Experimental Value | | PAW PBE | | PAW PW91 | | PAW LDA | |
|--------|--------------------|---|---------|---|----------|---|---------|---|
| | $a_{lat}$ (Å) | $B_o$ (GPa) | $a_{lat}$ (Å) | $B_o$ (GPa) | $a_{lat}$ (Å) | $B_o$ (GPa) | $a_{lat}$ (Å) | $B_o$ (GPa) |
| Ni | 3.524[a] | 180[a] | 3.523 | 193.635 | 3.52 | 196.743 | 3.426 | 252.468 |
| Fe | 2.866[a] | 170[a] | 2.834 | 204.0 | 2.827 | 199.095 | 2.747 | 252.234 |
| Cr | 2.910[a] | 160[a] | 2.855 | 177.235 | 2.841 | 212.345 | 2.779 | 305.626 |

[a]Reference 32

**Table 2.** Vacancy formation energies for Ni, Fe and Cr are calculated using PAW (PBE, PW91, LDA) pseudopotentials by DFT. Calculated values are compared with experimental data as well as other computed data.

| Metal | Experimental Value (eV) | Present Work PAW pseudopotential | | | Computed data by others | |
|---|---|---|---|---|---|---|
| | | PBE | PW91 | LDA | | |
| Ni | 1.78[a] 1.8[b] 1.79[d] | 1.42 | 1.37 | 1.64 | 1.77[e] 1.62[f] 1.37[f] | FP LMTO DFT (LDA) DFT (GGA) |
| Fe | 2±0.2[c] | 2.18 | 2.18 | 2.26 | 1.95[c] 1.93-2.07[c] 2.07[c] 2.12[g] 1.93[g] | VASP PW PWSCF PW SIESTA VASP PAW GGA VASP USPP GGA |
| Cr | 2.27[d] 2.0[b] | 2.30 | 2.66 | 2.81 | 2.86[e] 2.81[g] 2.81[g] | FP LMTO VASP PAW with AFM configuration VASP USPP with AFM configuration |

[a]Reference 33
[b]Reference 34
[c]Reference 35
[d]Reference 36
[e]Reference 37
[f]Reference 38
[g]Reference 39

**Table 3.** The computed DFT values of equilibrium lattice parameter, bulk modulus, vacancy formation energy, corrected vacancy formation energy are calculated using PAW PBE pseudopotential. The numbers are compared with experimental values [15] as well as data as calculated by Mattsson *et al.*[11]

| Metal | Electrons per atom | PP used | Exposed area due to monovacancy ($\text{Å}^2$) | Average bulk electron density ($e\text{Å}^{-3}$) | $\Delta\rho$ ($e\text{Å}^{-3}$) | Correction/area (MATT) ($eV/\text{Å}^2$) | Correction/area (CW) ($eV/\text{Å}^2$) | Total correction (MATT) (eV) | Total correction (CW) (eV) | $E_f^v$ (uncorrected) (eV) | $E_f^v$ (MATT) (corrected) (eV) | $E_f^v$ (CW) (corrected) (eV) | $E_f^v$ (Mattsson's work)[a] (eV) | $E_f^v$ (Expt.)[b] (eV) |
|---|---|---|---|---|---|---|---|---|---|---|---|---|---|---|
| Al | 3 | PAW PBE | 18.103 | 0.182 | 0.0958 | 0.0084 | 0.0044 | 0.152 | 0.080 | 0.62 | 0.77 | 0.70 | 0.78[a] | 0.68±0.03[b] |

[a]Reference 11
[b]Reference 15

**Table 4.** Corrected values of vacancy formation energies are compared with experimental values. Corrected values using average bulk valence electron density of the material and using $\Delta\rho$ are tabulated in columns labeled as "MATT" and "CW" respectively.

| Metal | Experimental value (eV) | PBE | | | PW91 | | | LDA | | |
|---|---|---|---|---|---|---|---|---|---|---|
| | | Without correction (eV) | With correction (MATT) (eV) | With correction (CW) (eV) | Without correction (eV) | With correction (MATT) (eV) | With correction (CW) (eV) | Without correction (eV) | With correction (MATT) (eV) | With correction (CW) (eV) |
| Ni | 1.78[a] 1.8[b] 1.79[d] | 1.42 | 1.95 | 1.58 | 1.37 | 2.07 | 1.58 | 1.64 | 1.85 | 1.70 |
| Fe | 2±0.2[c] | 2.18 | 2.58 | 2.36 | 2.18 | 2.71 | 2.41 | 2.26 | 2.42 | 2.32 |
| Cr | 2.27[d] 2.0[b] | 2.30 | 2.68 | 2.43 | 2.66 | 3.07 | 2.87 | 2.81 | 2.93 | 2.90 |

[a]Reference 33, [b]Reference 34, [c]Reference 35, [d]Reference 36

**Supplementary Information:**

The computed values of exposed surface area as well as corresponding surface corrections are calculated. Wigner-Seitz radius ($r_s$) are calculated using Equation 2. Values using average bulk valence electron density of the material and using $\Delta\rho$ are tabulated in columns labeled as "MATT" and "CW" respectively.

| Metal | Electrons per atom ($n_e$) | XC functional | Wigner-Seitz radius, $r_s$ (Å) | Exposed area due to monovacancy (Å²) | Average bulk electron density (eÅ⁻³) | $\Delta\rho$ (eÅ⁻³) | Correction/area (MATT) (eV/Å²) | Correction/area (CW) (eV/Å²) | Total correction (MATT) (eV) | Total correction (CW) (eV) |
|---|---|---|---|---|---|---|---|---|---|---|
| Ni | 10 | PBE | 0.639 | 13.772 | 0.914 | 0.252 | 0.0383 | 0.0116 | 0.527 | 0.158 |
| | | PW91 | 0.638 | 13.747 | 0.918 | 0.247 | 0.0507 | 0.0150 | 0.697 | 0.206 |
| | | LDA | 0.621 | 13.018 | 0.995 | 0.254 | 0.0160 | 0.0046 | 0.208 | 0.060 |
| Fe | 8 | PBE | 0.698 | 13.367 | 0.702 | 0.288 | 0.0301 | 0.0131 | 0.402 | 0.175 |
| | | PW91 | 0.696 | 13.295 | 0.708 | 0.291 | 0.0401 | 0.0175 | 0.533 | 0.233 |
| | | LDA | 0.676 | 12.558 | 0.772 | 0.275 | 0.0127 | 0.0050 | 0.159 | 0.063 |
| Cr | 6 | PBE | 0.773 | 13.558 | 0.516 | 0.200 | 0.0277 | 0.0093 | 0.376 | 0.126 |
| | | PW91 | 0.770 | 13.430 | 0.524 | 0.257 | 0.0303 | 0.0156 | 0.407 | 0.210 |
| | | LDA | 0.753 | 12.852 | 0.559 | 0.419 | 0.0095 | 0.0073 | 0.122 | 0.094 |